\documentclass[aps,twocolumn]{revtex4}

\usepackage{graphicx}
\begin{document}

\title{Conformational gel analysis and graphics: Measurement
of side chain rotational isomer populations by NMR and
molecular mechanics}
\author{Christopher Haydock}
\email{haydock@appliednewscience.com}
\affiliation{Applied New Science LLC, Rochester, Minnesota 55901, USA}

\begin{abstract}
Conformational gel analysis and graphics systematically identifies and
evaluates plausible alternatives to the side chain conformations found
by conventional peptide or protein structure determination methods.
The proposed analysis determines the populations of side chain rotational
isomers and the probability distribution of these populations.
The following steps are repeated for each side chain of a peptide or protein:
first, extract the local molecular mechanics of side chain rotational
isomerization from a single representative global conformation; second, expand
the predominant set of rotational isomers to include all probable rotational
isomers down to those that constitute just a small percentage of the
population; and third, evaluate the constraints vicinal coupling constants
and NOESY cross relaxation rates place on rotational isomer populations.
In this article we apply conformational gel analysis to the cobalt
glycyl-leucine dipeptide and detail the steps necessary to generalize the
analysis to other amino acid side chains in other peptides and proteins.
For a side chain buried within a protein interior, it is noteworthy that the
set of probable rotational isomers may contain one or more rotational isomers
that are not identified by conventional NMR structure determination methods.
In cases such as this the conformational gel graphics fully accounts
for the interplay of molecular mechanics and NMR data constraints on the
population estimates.  The analysis is particularly suited to identifying side
chain rotational isomers that constitute a small percentage of the population,
but nevertheless might be structurally and functionally very significant.
\end{abstract}

\maketitle

\section{INTRODUCTION}

Simple changes in functional groups can often make multiple
order-of-magnitude changes in biological activity \cite{Gilman05}.
This suggests that protein conformations populated at the 10\%,
1.0\%, or even 0.1\% level may play a very significant role
in function.

The most common NMR protein
structure determination methods generate perhaps a few dozen
complete protein structures \cite{Guntert05}.
The structures in this ensemble are independently fit to the data
and each final structure should fit the data about equally well.
The optimization of a multiconformer model differs sharply
from these standard fitting methods because the measure
of goodness-of-fit is not defined for any single
structure in the ensemble, but depends simultaneously
on the entire ensemble of structures.  In contrast to the standard
methods, no single structure in the final multiconformer ensemble
need be a particularly good fit to the data \cite{Brunger97a}.
In favorable cases multiconformer models can give some indication
of the conformational variability of proteins in solution \cite{Brunger97b}.
For example, particular secondary structure elements,
loops, or even side chains of the multiconformer model might have
a larger RMS deviation indicating variability of these parts in solution.
Information about local variability thus comes from a global fitting procedure.
Whether conformational variability is assessed by conventional
methods or multiconformer models, the essential idea is to narrow
down the vast global conformational space of a protein by applying the
constraints of real data.  This point also marks one of the key
differences of conformational gel analysis because conformational
gel analysis identifies local conformations based solely on molecular
mechanics and a single representative global conformation previously
determined from current or preexisting NMR data or even crystallographic data.
Instead of applying NMR data to pick out new global conformations,
the NMR data is analyzed to determine the extent to which
it constrains the populations of the local conformations identified
by molecular mechanics.

Detailed information about local conformations
is often available from molecular mechanics.
In the case of protein side chain rotational isomerization,
good estimates of the position and shape of potential energy wells are known
and approximate depths of these wells are also known \cite{Haydock93}.
Even in cases where such information might be exploited to reduce or
eliminate the conformational search problem, its potential use is 
overlooked or even dismissed, apparently because the experimental
data is judged more reliable than
the molecular mechanics results \cite{Landis95},
the molecular mechanics geometries and energies may be judged 
globally coupled to such an extent that local conformational information
can not be separated out \cite{Kozerski97,James98,Pearlman96},
or perhaps because the molecular mechanics models are not
readily available \cite{Schmidt97,Dzakula96,Hu97}.

The cobalt glycyl-leucine
dipeptide \footnote{Barium[glycyl-L-leucinatonitrocobalt(III)].}
NMR data analyzed in this article was previously analyzed
in a more comprehensive and perhaps less accessible manner \cite{Haydock00}.
The present analysis of this data is not designed
to extend the knowledge of the cobalt dipeptide system,
but rather to be readily generalizable to proteins.
The cobalt dipeptide analysis is particularly suited
to generalization to proteins because 
the cobalt chelate ring system fixes the dipeptide
backbone in a definite conformation and the simplifying assumption
of a single backbone conformation also applies to the analysis of proteins.

\section{CONFORMATIONAL GEL ANALYSIS AND GRAPHICS}

\begin{figure}
\resizebox{8.0cm}{!}{\includegraphics{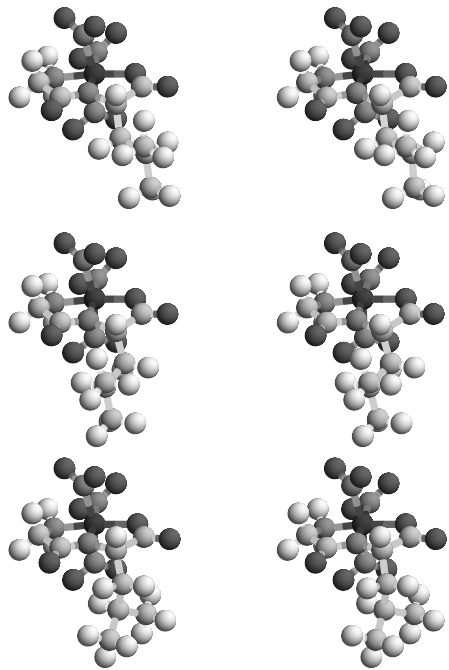}}
\caption{Stereo views of the predominant rotational isomers
of the leucine side chain of the cobalt glycyl-leucine dipeptide:
\textit{top}, trans gauche$^+$; \textit{middle}, gauche$^-$ trans;
and \textit{bottom}, gauche$^+$ gauche$^+$.
The atom gray scale tones are: \textit{white}, hydrogen;
\textit{light gray}, carbon; \textit{medium gray}, nitrogen;
\textit{dark gray}, oxygen; \textit{black}, cobalt.
The leucine side chain projects outward towards the viewer and the
three chelated nitro groups are visible below, behind, and above
the cobalt dipeptide ring system.}
\label{fig1:dipeptide_molecule}
\end{figure}

Conformational gel analysis and graphics provides
detailed molecular mechanics population estimates and assesses
the constraints that NMR data places on these population estimates.
Much of this information is expressed in the form of gel graphics.
Though gels are widely applied to the separation, characterization,
and identification of all types of biomolecules
and their developed images are widely seen in the biochemical literature,
the gel graphics employed here are entirely computer generated
and their interpretation is in many ways novel.
The basic facts about their interpretation are perhaps
most easily explained by showing their connection
to molecular mechanics energy plots.
We will make this connection in the opening paragraphs
of this section in a highly simplified analysis
of the three predominant rotational isomers
of the leucine side chain of the cobalt dipeptide
(Figure \ref{fig1:dipeptide_molecule}) and then
move on to a detailed explanation of conformational gel analysis
and graphics in the three numbered subsections of this section.

\begin{figure}
\resizebox{8.0cm}{!}{\includegraphics{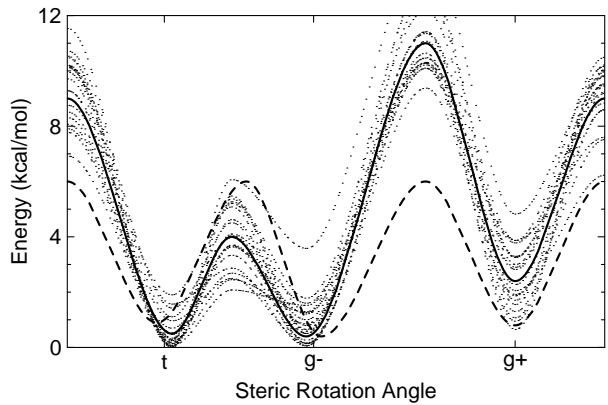}}
\caption{Molecular mechanics energy schematic for the three
predominant leucine side chain rotational isomers.
Isomerization energies are plotted as a function of the
$\chi^1$ torsion angle for rotation about the leucine side chain
$\alpha$ to $\beta$-carbon bond.
The horizontal axis labels refer to only the $\chi^1$
torsion angle of each rotational isomer.
\textit{Solid}, molecular mechanics energy function;
\textit{dashed}, NMR experimental data energy function;
\textit{dotted}, energy function distribution from estimated
molecular mechanics errors.  
This energy function distribution displays essentially
the same information as a gel graphic.}
\label{fig2:reaction_coordinate}
\end{figure}

\begin{figure}
\resizebox{8.0cm}{!}{\includegraphics{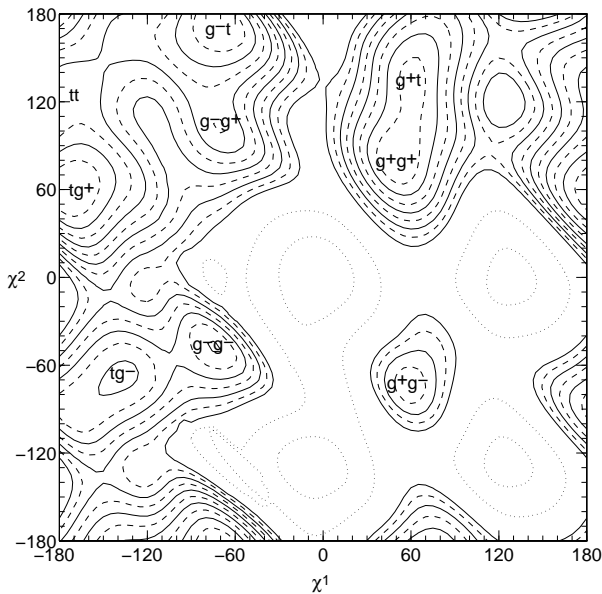}}
\caption{Molecular mechanics energy map for rotational
isomerization of cobalt dipeptide leucine side chain.
Contour levels are \textit{dashed}, 1, 3, 5, 7, 9;
\textit{solid}, 2, 4, 6, 8, 10; \textit{dotted}, 15, 20 kcal/mol.
Zero corresponds to $-$39.4 kcal/mol.
The nine rotational isomers are labeled at the position
of their energy well minima.}
\label{fig3:contour_plot_x1x2}
\end{figure}

A gel graphic can convey the uncertainty of rotational isomer populations
obtained either by molecular mechanics calculations or by fitting NMR data.
The distinction between calculated or fit populations and the
uncertainties of these populations parallels that between
a simple energy function and an energy function distribution
(Figure \ref{fig2:reaction_coordinate}).  In this example
all the energy functions give energies for rotation about
the leucine side chain $\alpha$ to $\beta$-carbon bond.
The three troughs of each sinusoidal function are the energy wells
of the three predominant rotational isomers and the three crests
are the energy barriers to interconversion.
Note that the molecular mechanics energy function
(Figure \ref{fig2:reaction_coordinate}, \textit{solid})
can be calculated from the full $\chi^1 \times \chi^2$ energy map
(Figure \ref{fig3:contour_plot_x1x2}) by rotating the $\chi^1$
torsion angle and as appropriate adjusting the $\chi^2$ torsion angle
in such a way as to pass over the energy barriers separating
the three predominant rotational isomers, see methods.

The molecular mechanics energy function is calculated from
an empirical energy function, which has many parameters,
such as bond length and bond angle equilibrium values,
torsion angle phases and multiplicities, force constants,
atomic partial charges, and Lennard-Jones constants \cite{MacKerell98}.
These parameters are fit to theoretical and experimental
data for model compounds.  Errors are introduced by this data,
by the necessary simplicity of an empirical energy function,
and by the need to transfer parameters from the simple model compounds
to a larger molecule of interest, such as the cobalt dipeptide.
The $\chi^1 \times \chi^2$ energy map on which this example
is based is calculated by energy minimization with the $\chi^1$
and $\chi^2$ torsion angles constrained and without any explicit
solvent.  This makes for quick calculation, but introduces
further errors because there is no averaging over other cobalt
dipeptide internal degrees of freedom nor over any solvent degrees of freedom.
The molecular mechanics energy function
(Figure \ref{fig2:reaction_coordinate}, \textit{solid})
is a best overall estimate of the rotational isomerization energy
and each energy function in the distribution of energy functions
(Figure \ref{fig2:reaction_coordinate}, \textit{dotted})
represents a possible deviation from this best estimate
due to all the above errors.

\begin{figure}
\resizebox{8.0cm}{!}{\includegraphics{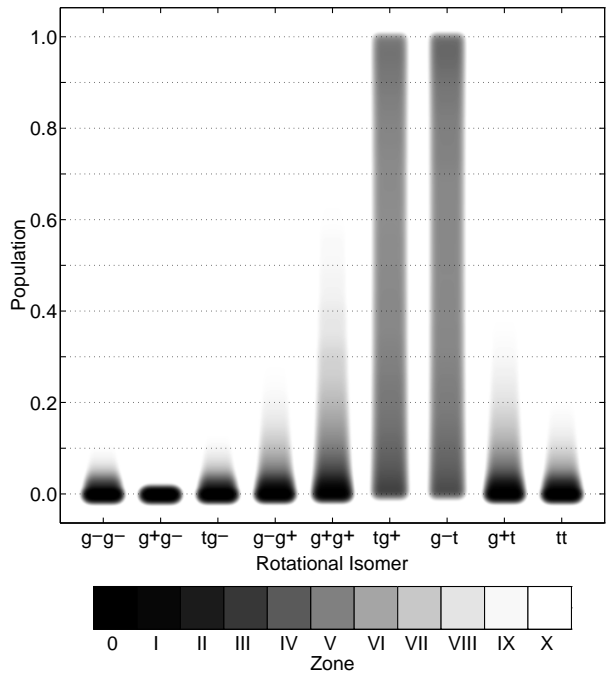}}
\caption{Molecular mechanics gel graphic
for the cobalt dipeptide leucine side chain.
The gel graphic is calculated from the molecular mechanics energy map
in the absence of the experimental data and identifies the probable set
of rotational isomers for further conformational gel analysis.
The population probability distributions shown in the gel graphic
are calculated for an uncertainty of the map energy well depths
of $\pm$1.0 kcal/mol and a temperature of 300 K.
Each gray scale step of the stepwedge bar corresponds
to a two-fold change in probability density.}
\label{fig4:map_gel_graphic}
\end{figure}

The Boltzmann factor establishes a relation
between the energy and the population of a state.
The connection between rotational isomer population uncertainties
and an energy function distribution follows from this relation.
For example the energies at the trans, gauche$^-$, and gauche$^+$ well
minima of the molecular mechanics energy function are
0.5, 0.4, and 2.4 kcal/mol
(Figure \ref{fig2:reaction_coordinate}, \textit{solid})
and at 300 degrees K the Boltzmann factors for these energies give
populations of 45, 53, and 2\% for the three predominant rotational isomers.
It is also important to note that though it is meaningful
to talk about the energy of rotational isomerization of a single
molecule, the population must refer to an ensemble average
over many molecules or at least to a time average for a single molecule.
Even though there is an ensemble of molecules there is no uncertainty
in the populations predicted by the simple energy function.
The population probability distributions
(Figure \ref{fig4:map_gel_graphic}, lanes 5--7)
only arise from the distribution of energy functions
(Figure \ref{fig2:reaction_coordinate}, \textit{dotted}).
Note that probability distributions for the trans, gauche$^-$, and
gauche$^+$ predominant rotational isomers in this example are only
slightly affected by the other rotational isomers in the probable set
(Figure \ref{fig4:map_gel_graphic}, lanes 1--4, 8, and 9)
because the probability distributions of all these other
rotational isomers favor very low populations.
Each lane in the molecular mechanics gel graphic is itself
a probability distribution for one rotational isomer
and the total probability for each rotational isomer
is always normalized to one.
If the uncertainties in the well depths are all zero,
then the population of each rotational isomer is precisely determined.
In contrast, nonzero uncertainties in the well depths
give population probability distributions, which
appear on the gel graphic as either broadened or extended bands.
The molecular mechanics gel graphic gives explicit visual representation
to the errors inherent in molecular mechanics.

\begin{figure}
\resizebox{8.0cm}{!}{\includegraphics{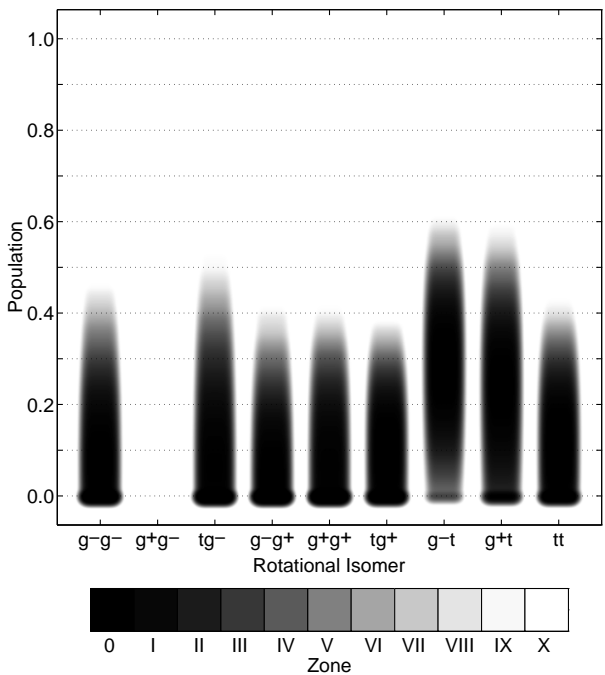}}
\caption{Conformational gel graphic for the entire probable set
of rotational isomers of the cobalt dipeptide leucine side chain.
The gel graphic visually portrays the extent to which the NMR data
constrains the populations of the probable set of isomers.
The population probability distributions shown in the gel graphic
are constructed by repeated fitting of the rotational isomer
populations to Monte Carlo NMR data.
Each gray scale step of the stepwedge bar corresponds
to a two-fold change in probability density.}
\label{fig5:most_likely}
\end{figure}

\begin{figure}
\resizebox{8.0cm}{!}{\includegraphics{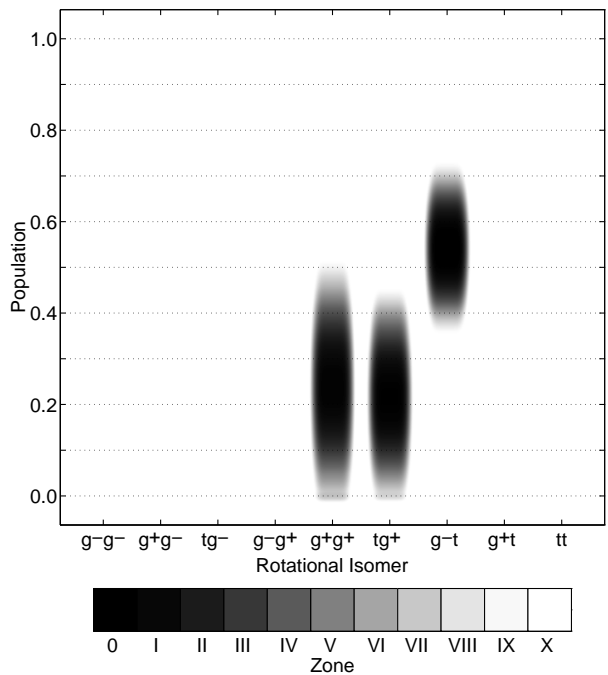}}
\caption{Conformational gel graphic for the predominant set
of rotational isomers of the cobalt dipeptide leucine side chain.
The restriction of the probable set of rotational isomers
to those that molecular mechanics suggest are predominant
clearly reduces the uncertainty in the populations.
For a protein side chain such a restriction is not always desirable
because less conspicuous rotational isomers might be structurally
and functionally very significant.}
\label{fig6:pdb_g_plus}
\end{figure}

The interpretation of conformational gel graphics
(Figures \ref{fig5:most_likely} and \ref{fig6:pdb_g_plus})
is in many respects similar to that of a molecular mechanics
gel graphic as described above.  Again we will focus on the
populations estimates for the three predominant rotational
isomers (Figure \ref{fig6:pdb_g_plus}, lanes 5--7)
and leave detailed consideration of the entire probable set
of rotational isomers to a later subsection.
If a model with only the trans, gauche$^-$, and gauche$^+$ predominant
rotational isomers is fit to the NMR data for the cobalt dipeptide,
the predicted populations of these rotational isomers are 21, 54, and 25\%.
Though it is natural to fit the NMR data by adjusting
rotational isomer populations rather than well energies,
these fit populations may be converted to energies by inverting
the Boltzmann factor procedure described in the previous paragraph.
At 300 degrees K the fit populations are equivalent to a simple energy
function with trans, gauche$^-$, and gauche$^+$ energy well minima
at 0.9, 0.4, and 0.8 kcal/mol
(Figure \ref{fig2:reaction_coordinate}, \textit{dashed}).
The standard Monte Carlo method for estimating the distribution
of the rotational isomer population estimates assumes that
the estimated populations are the true populations,
generates a large number of simulated NMR data sets,
and fits these simulated NMR data sets to genterate the population
distribution, see methods.
The gel graphic (Figure \ref{fig6:pdb_g_plus}, lanes 5--7)
displays this population distribution.
By the inverted Boltzmann factor procedure
the population distribution could be converted into an energy
function distribution and plotted in an energy schematic.
In the same way that the molecular mechanics gel graphic
(Figure \ref{fig4:map_gel_graphic}, lanes 5--7)
has an equivalent energy function distribution
(Figure \ref{fig2:reaction_coordinate}, \textit{dotted})
so also does a conformational gel graphic
(Figure \ref{fig6:pdb_g_plus}, lanes 5--7)
have an equivalent energy function distribution (not shown).
Only the origins of the distributions differ.
The molecular mechanics energy function distribution
is generated directly by applying Monte Carlo energy errors
to a simple molecular mechanics energy function,
but the energy function distribution equivalent to a conformational
gel is generated from the underlying rotational isomer population
distribution, which is in turn indirectly generated from
Monte Carlo NMR data errors as outlined above.
In short, the conformational gel graphic gives explicit visual
representation to the errors inherent in the NMR data.

We now turn to a detailed explanation of the conformational gel
analysis and graphics of the cobalt dipeptide and along the way
indicate how this analysis can be generalized to proteins.

\subsubsection{Extract local molecular mechanics 
from a single representative global conformation}

In this work we confine our attention to the
rotational isomerization of peptide and protein side chains
as examples of local molecular mechanics.
The cobalt dipeptide energy map for the rotational isomerization
of the leucine side chain (Figure \ref{fig3:contour_plot_x1x2})
is a very simple example of local molecular mechanics.
To calculate this map the leucine side chain torsion
angles are constrained to values on a 5 degree grid over the
$\chi^1 \times \chi^2$ torsion space and the entire dipeptide
structure is energy minimized \cite{Haydock00}.
To extract local molecular mechanics of proteins
local backbone flexibility and local side chain interactions
must be carefully controlled \cite{Haydock93}.
A single representative global conformation can be routinely
extracted from an ensemble of NMR protein structures by averaging
and constrained minimization \cite{Guntert05}.

Though only the side chain conformation
at an energy minimum of a well is required for the gel analysis,
the entire map is useful for automated identification
of the energy minima and is absolutely essential for correctly
controlling the local backbone flexibility and neighbor
side chain interactions.  To control the backbone flexibility
the entire protein backbone is fixed except for
backbone segments of two or at most three amino acids.
Essentially, the number and length of these free atom segments
must be sufficient to accurately determine the position
of the energy well minima, but this number and these lengths
must not be so unduly generous as
to make energy minimizations unnecessarily expensive
or as to obscure the energy map with transitions of nonlocal
backbone degrees of freedom.  The energy map is very effective
tool for eliminating these nonlocal transitions because
they show up as discontinuities of the energy surface.
If a map has these discontinuities then it must be recomputed
with reduced backbone flexibility.  The accuracy of the energy
map can be judged by comparing the position, shape and depth
of energy wells of maps computed at two or three different
levels of backbone flexibility.

The effects of neighbor side chain interactions are assessed
by truncating neighbor side chains at the $\beta$-carbon.
By comparing the shape and position of energy wells of maps
calculated with and without neighbor side chain truncation
it is possible to gauge the extent to which particular
interactions influence particular side chain rotational isomers.
In rare cases a neighbor side chain interaction may
be judged to preclude a particular energy well at any potentially
interesting level of population.  More commonly a neighbor side chain
interaction will simply increase the uncertainty of a target well's energy
depth because the uncertainty of the interaction strengths
of the target well with all of the neighbor side chain wells
and the uncertainty in the energy depth of all the neighbor wells
must be folded into the uncertainty of the target well's depth.

The energy map for the leucine side chain
of the cobalt dipeptide (Figure \ref{fig3:contour_plot_x1x2})
is similar to maps computed for side chains in a variety of protein
environments.  One important similarity is that
during calculation of the side chain energy map
the cobalt dipeptide backbone is fixed in a definite conformation
by the cobalt chelate ring system.  This parallels the approach
to calculating a protein side chain energy map described above,
where the protein backbone is fixed in the same conformation as
found in the single representative global conformation.
The cobalt dipeptide and protein maps both give the
local molecular mechanics of side chain rotational isomerization.
The leucine map of the cobalt dipeptide is also similar
to protein side chain maps in that the total number of wells
present in the map is the same or close to the number of ideally possible
rotational isomers, for example, nine in the case of a leucine side chain.
This is true even for a side chain buried within a protein because
neighbor side chains are usually truncated at the $\beta$-carbon
in order not to exclude target side chain conformations that
interact unfavorably with the neighbor side chain conformation
that happens to exist in the single representative global conformation
of the protein.  As already mentioned the uncertainty of the energy
depth of such target side chain wells is substantially increased.
The positions of the well energy minima on the energy map
for the leucine side chain of the cobalt dipeptide differ
from their ideal positions by from about 10 degrees (trans gauche$^+$
rotational isomer) to almost 60 degrees (trans trans rotational isomer).
The average departure from ideal positions tends to increase for side
chains buried within proteins.

\subsubsection{Expand the predominant set to include
all probable rotational isomers}

The single representative global conformation determined
by conventional methods already gives a preferred conformation
for each side chain of a protein.
For each side chain it is usually possible to adjoin the
preferred rotational isomer with one or two others and form
a predominant set of rotational isomers that account for say
90\% of the population of the side chains rotational isomers.
The side chain rotational isomers included in this predominant set
might be present in an ensemble of conformations
determined by conventional NMR, would probably have low energy wells
in the molecular mechanics energy map,
and might possibly be suggested by rotamer preference libraries
complied from the protein data bank.
There are several different scenarios that may arise
and these are best illustrated by referring to the cobalt dipeptide
molecular mechanics gel graphic (Figure \ref{fig4:map_gel_graphic}).

The molecular mechanics gel graphic plays an important role both
in identifying the predominant set of rotational isomers and in expanding
this set to make the set of all probable rotational isomers.
As discussed in the opening paragraphs of this section
this gel graphic differs from the energy map in that it takes
into account not only the energy depth of each rotational isomer's
well, but also the uncertainty of the energy well depths.
A comparison of cobalt dipeptide gel graphic and gel graphics computed
for side chains in a variety of protein environments follows along much
the same lines as the comparisons between energy maps in the previous section.
Like a protein energy map, the resulting protein molecular mechanics gel
reflects the local molecular mechanics of the target side chain.
Just as the protein energy map usually has energy wells corresponding
to each of the ideally possible rotational isomers,
the resulting protein molecular mechanics gel has the same corresponding lanes.
Unlike the well depth uncertainties of the cobalt dipeptide molecular
mechanics gel, which are all equal because there are no neighbor
side chains, the uncertainties of a protein
gel would be significantly larger for rotational isomers
that interact strongly with neighbor side chains.
It is difficult to draw
reliable conclusions from the molecular mechanics energy map
without taking these uncertainties into account.

From the cobalt dipeptide molecular mechanics gel graphic
(Figure \ref{fig4:map_gel_graphic}) the predominant rotational isomers
of the leucine side chain are trans gauche$^+$ and gauche$^-$ trans
and possibly gauche$^+$ gauche$^+$.
Analysis of NMR data suggests that these first
two rotational isomers are the most populated and that the third
may also make a significant contribution \cite{Haydock00}.
This agreement between experiment and molecular mechanics
comes about because the experimental analysis is designed to reproduce
the rotational isomer preferences observed in the protein data bank
\cite{Dunbrack94}, which not coincidently are much the same as the
relative rotational isomer stabilities predicted by molecular mechanics.
We expect that for many protein side chains the same set of predominant
rotational isomers will be given by the molecular mechanics gel graphic
and the ensemble of conformations determined by conventional NMR,
but for slightly different reasons than those just mentioned
for the cobalt dipeptide leucine side chain.
Perhaps the most important reason that the molecular mechanics energy map
is likely to predict the same predominant rotational isomers
is that it is computed with the backbone fixed in the same conformation
as found in the single representative global conformation.
The side chain energy map is thus likely to favor the rotational isomers
populated in the ensemble of conventional NMR conformations
because the map is computed with the backbone fixed in a conformation
that is representative of this very same ensemble.

The set of all probable rotational isomers includes all the predominant
rotational isomers as well as those that make smaller contributions
to the population down to contributions as small as
perhaps a few tenths of a percent.  The molecular mechanics
gel graphic (Figure \ref{fig4:map_gel_graphic}) for the cobalt dipeptide
leucine side chain suggests that all rotational isomers except
gauche$^+$ gauche$^-$ are in the probable set.  This is shown visually
by a single band at zero population and a complete lack of any
upwards extension of this band.
As mentioned earlier in this section the molecular mechanics
gel graphics of protein side chains have lanes corresponding
to most of the ideally possible rotational isomers.
Even for a side chain buried in a protein interior
a good fraction of these will at least be in the probable set
of rotational isomers.  Compared to the leucine side chain of
the cobalt dipeptide the distinction between predominant and probable
sets of rotational isomers may not be as clear cut because
of the increased energy uncertainty of rotational isomers
that strongly interact with a neighbor side chain.
A large uncertainty shows up on the molecular mechanics gel graphic
as a lane with bands at population zero and one and a much
fainter extensions stretching between the these two extremes.
Rotational isomers with such a large molecular mechanics
energy uncertainty should certainly be included in the probable set
and may even be sufficiently populated to be included in the
predominant set \cite{Haydock93}.

\subsubsection{Evaluate the constraints vicinal coupling constants and
NOESY cross relaxation rates place on rotational isomer populations}

Thus far we have described the molecular mechanics of the leucine
side chain of the cobalt dipeptide, the selection of predominant
and probable rotational isomers, and how this can be generalized
to side chains of proteins.
The key point about the generalization to proteins is that
the molecular mechanics remains local.
By local it is meant that the molecular mechanics model depends
on the single representative global conformation, which
is always readily available from the preexisting NMR structure.
The side chain molecular mechanics does not depend
on a multiconformer model, which would in some way require
the simultaneous solution of all the side chain rotational
isomer populations.  In this section the molecular mechanics
model is fit to the NMR data to evaluate the populations
of the predominant and the probable rotational isomers.
This can be generalized to proteins by exploiting the locality
of both the molecular mechanics model and NMR data to carry out the
evaluation independently for each side chain.

The conformational gel graphic for the probable set of rotational
isomers of the cobalt dipeptide leucine side chain
(Figure \ref{fig5:most_likely}) shows that the experimental data,
which consists of eight vicinal coupling constants and ten NOESY
cross relaxation rates \cite{Haydock00}, places little constraint
on the populations of these eight isomers.
This is indicated by the intense bands extending from zero population
up to thirty to fifty percent population for each of the
rotational isomers in the probable set.
A comparison of the molecular mechanics and conformational gel graphics
(Figures \ref{fig4:map_gel_graphic} and \ref{fig5:most_likely})
gives a striking graphical portrayal of the dramatic
variation in the usefulness of molecular mechanics and NMR data 
for determining rotational isomer populations.
Apparently, the populations of most of the cobalt dipeptide
side chain rotational isomers are best determined either
by the NMR data alone or by the molecular mechanics calculations alone.
Except for the three more extended lanes
(gauche$^+$ gauche$^+$, trans gauche$^+$, and gauche$^-$ trans)
near the middle of the molecular mechanics gel graphic
(Figure \ref{fig4:map_gel_graphic})
all the rest of the rotational isomers have bands at zero
population that at most have a relatively small upwards extension.
These rotational isomer populations are better
determined by molecular mechanics.
In contrast the trans gauche$^+$ and gauche$^-$ trans rotational isomers
have bands stretching all the way from population zero up to one
in the molecular mechanics gel graphic (Figure \ref{fig4:map_gel_graphic})
as compared to much less extended bands in the conformational
gel graphic (Figure \ref{fig5:most_likely}).
These rotational isomer populations are better
determined by fitting the NMR data.
The similarity of the gauche$^+$ gauche$^+$ rotational isomer bands
in both figures suggests both molecular mechanics and NMR data must
be taken into account to determine the population of this rotational isomer.
All these conclusions are born out by more detailed analysis \cite{Haydock00}.

Clearly it is desirable to reduce the size of the probable
set of rotational isomers to the point that the NMR data does
place significant constraints on the populations.
In the case of the cobalt dipeptide this point is reached
when the probable set is reduced to the three rotational isomers
of the predominant set defined in the previous section.
The conformational gel graphic for the predominant set
(Figure \ref{fig6:pdb_g_plus}) displays population errors
that are somewhat to considerably smaller than the predicted populations
of the rotational isomers.  This is displayed visually by the
modest to large distance from the zero population horizontal
grid line and to the beginning of the high density region of the bands.
Note that the bands extend nearly three standard deviations above and
below the mean, but the high density region extends only about two
standard deviations in each direction.
Comparing the conformational gel graphics for the probable
and the predominant sets of rotational isomers
(Figures \ref{fig5:most_likely} and \ref{fig6:pdb_g_plus})
a particularly striking improvement is seen in the significance
of the population estimate of the gauche$^-$ trans rotational isomer.

For any protein side chain it is in principle possible
to define a predominant set of rotational isomers that
is just small enough to yield significant population estimates.
The practical difficulty with this is that the molecular mechanics
does not always give reliable ordering of the energy well depths
because of the sources of error discussed in the previous section.
As the size of the probable set is reduced it will not always
be clear which rotational isomers to include or exclude.
All three gel graphics (Figures \ref{fig4:map_gel_graphic},
\ref{fig5:most_likely}, and \ref{fig6:pdb_g_plus}) must
be considered together to obtain a complete picture
of the rotational isomer populations
that fully accounts for the interplay of molecular mechanics and NMR data.
A still more detailed analysis should also consider
measurability and over-fitting of rotational isomer populations
\cite{Haydock00}, but an easily accessible description of the
application of these concepts is beyond the scope of this article.

\section{CONCLUSIONS}

This work makes new theoretical predictions of interest to a broad
range of chemists studying the structure and function of proteins
or other complex molecules.
Particularly important is the prediction that
the local molecular mechanics of protein side chains can be extracted from
a single representative global conformation determined by conventional methods.
The local molecular mechanics can identify low population though potentially
functional rotational isomers of buried protein side chains.
Conformational gel analysis and graphics is an important new tool
for display and understanding of conformational population estimates
and of the sources and level of errors in these estimates.
By helping us see more clearly the extent of both our knowledge
and our ignorance we hope to fuel the demand and inspire and guide
the development of more powerful NMR instruments and analysis methods.

\section{METHODS}

Detailed descriptions as well as working computer input files
for calculating molecular mechanics energy maps, fitting NMR data,
and generating gel graphics, have been previously published
\cite{Haydock93,Haydock00}.
Briefly, custom topology and parameter input files were created
and $\chi^1 \times \chi^2$ energy maps for the leucine side chain
of the cobalt dipeptide were calculated with the CHARMM
molecular mechanics program \cite{Brooks83}.
Based on the positions of the energy well minima
nine energy minimized rotational isomers were prepared and
interatomic distances and torsion angles for modeling
cross relaxation rates and vicinal coupling constants were
output with the CHARMM correlation and time series analysis command.
The optimization design matrix was obtained by a MATLAB function file that
input a list of NOESY cross relaxing protons and vicinally coupled spins,
read in the appropriate distance and angle data files output
by the analysis command,
calculated the cross relaxation rates and the vicinal coupling constants
for each rotational isomer, and normalized each of these observables
by a composite experimental and cross relaxation rate or
Karplus coefficient error.
Note that the original version of this MATLAB function file \cite{Haydock00}
was somewhat more complex than described here
because it was designed to examine the effects of intramolecular motions
by averaging over the molecular mechanics energy map.
The rotational isomer populations were fit by minimizing
the differences between the experimentally measured and predicted
observables subject to the constraints that the populations
were nonnegative and that their sum was one.
This linear least-squares with linear constraints problem
was solved as the equivalent quadratic programming problem \cite{Gill81}.
The probability density functions of the fit rotational isomer populations
were computed by the standard Monte Carlo recipe \cite{Press89}:
the experimental NMR observables were fit
to yield fit rotational isomer populations and fit NMR observables,
random errors were added to the fit NMR observables and
these simulated NMR observables were fit to give simulated fit populations,
these last steps were repeated may times to make the Monte Carlo
probability density functions of the populations.

The energy functions in the energy schematic were obtained by cubic
interpolation from the positions of the energy minima and maxima.
The derivatives of the interpolating functions were constrained
to zero at the positions of these minima and maxima.
Each energy function was shifted by an energy constant so that
at 300 degrees K the sum of the Boltzmann factors of the energy
minima equalled one.
The molecular mechanics energy function was obtained from
the molecular mechanics energy map by matching
the energies of the predominant rotational isomer minima
and the energies of the lowest energy barriers between
the predominant rotational isomers.
The horizontal positions of the energy function minima and maxim
were adjusted somewhat to make the maximum slopes of the energy
barriers about equal while still approximatly matching
the rotation of the $\chi^1$ torsion angle.
The horizontal axis of the energy schematic is labeled steric rotation
angle rather than $\chi^1$ torsion angle to reflect this approximation
and to emphasize the steric relationship between the side chain atoms
rather than the exact rotation angle.

The molecular mechanics gel graphic was generated from
the energy map by a simple Monte Carlo procedure.
Random energy errors were added to the rotational isomer well depths
and Boltzmann weighted populations were calculated and normalized,
these steps were repeated many times,
and the resulting large set of simulated rotational isomer populations
was histogramed to make Monte Carlo
probability density functions of the populations.

Monte Carlo probability density functions were
displayed as gel graphics, which were designed to visually indicate both
the discrete probability fraction at zero population and shape of the
continuous probability density over the range of population from zero and one.
This was accomplished by a simulated photographic process where the degree
of film overexposure indicates the probability fraction at zero population
and continuous gray tones represent the continuous part of the probability
density.  To simulate film overexposure at zero population and smooth
the lane edges along the continuous part of the probability distribution
a Gaussian blur filter was applied to the image so that
the typical probability density at zero population was still
considerably greater than that along the continuous part
of the probability distribution.
The pixel values of the blurred image were treated like scene
luminances \cite{Zalewski95} and converted into photographic
print densities with a characteristic curve \cite{Hunt95}
that had a maximum point-gamma of 1.5.
The print densities were linearly mapped into gray scale values
so that maximum printable density was somewhere along
the continuous part of the probability distribution.
A stepwedge bar of the 11 zones in the Zone System \cite{Adams95}
was added to the gel graphic as an aid to calibrating
the probability densities.

Vector PostScript molecular graphics were generated with the RasMol
program \footnote{Web site http://www.umass.edu/microbio/rasmol/ .}.

\section{ACKNOWLEDGMENT}
This work was financially supported by Franklyn G. Prendergast in the
Department of Biochemistry and Molecular Biology at the Mayo Clinic
in Rochester, Minnesota, USA.

\bibliography{gel}

\end{document}